# *Design, customization and implementation of energy simulation with 5E model in elementary classroom*

Sze Yee Lye[a], Loo Kang Wee[a], Yao Chie Kwek[b], Suriati Abas[b], Lee Yong Tay(Dr)[b]
lye_sze_yee@moe.edu.sg, wee_loo_kang@moe.edu.sg, kwek_yao_chie@moe.edu.sg,suriati_abas@moe.edu.sg, tay_lee_yong@moe.edu.sg
[a] Educational Technology Division, Ministry of Education (Singapore)
[b] Beacon Primary School (Singapore)

Corresponding author: Sze Yee Lye (lye_sze_yee@moe.edu.sg)

## Abstract

Science simulations are popular among educators as such simulations afford for multiple visual representation and interactivity. Despite the popularity and abundance on the internet, our literature review suggested little research has been conducted on the use of simulation in elementary school. Thus, an exploratory pilot case study was conducted to address this research gap. In this study, an open source energy simulation was remixed for use in elementary school targeted at the Grade 4 & 5 students as an after-school enrichment program. We proposed 3 stages: design, customization and implementation, to provide useful insights with the aim to allow other educators to conduct their own remixed simulation lessons. The simulation design principles (e.g., learning outcomes and colour coding) with the corresponding TPACK construct that emerged from the design and customization stages were reported. Such simulation design principles would be useful to interested educators and researchers who wish to adapt and use simulation or teach others how to remix simulation. Data from the multiple sources (e.g., field observations, surveys, design notes and existing simulations) indicated that students enjoyed learning with the remixed energy simulation.

## Keywords

Science simulation, Elementary school, Open source physics, Energy, TPACK

## Introduction

Science simulations are best described as “software programs that allow students to explore complex interactions among dynamic variables that model real-life situations” (Park, Lee, & Kim, 2009, p. 649). Like Rutten, van Joolingen, and van der Veen (2012), we view interactivity to be an essential feature of science simulations and this is what that makes them distinct from non-interactive based animations. Simulations are capable of accepting inputs and presenting the computational results in multiple representations like graphs or tables. Such interactivity and multiple representations afforded by simulation lends itself as a handy tool for inquiry-based learning, a common approach adopted by science educators (S. Chen, 2010). Moreover, a wealth of such simulations is freely available and accessible online (S. Chen, 2010). With their educational usefulness and availability, such simulations, inevitably, are becoming popular among Science educators.

Despite the pervasive use of simulations in schools and considerable research reporting on older students (Y. L. Chen, Hong, Sung, & Chang, 2011; Lamb & Annetta, 2013; Rutten et al., 2012; Scalise et al., 2011), there is paucity of research regarding the use of simulations in elementary school settings (Smetana & Bell, 2012). To address this gap, we report a study that explores the design, customization and implementation of energy simulation (Gallis, 2010) in an elementary classroom. We modified this open source roller coaster simulation created using Easy Java Simulation Toolkit developed by Esquembre (2012). Taking on the dual role of a designer and teacher, we modified the simulation for use to suit the elementary classroom and also implemented the lesson the based on the BCSE 5E model (Bybee et al., 2006).

## Literature Review

### Inquiry Learning with Science Simulation

Inquiry learning has been an essential part of Science education throughout the world. Students are situated in an inductive learning mode, in which deductions on science principles are made based on their experience with the instructional materials (de Jong, 2005). In this paper, we adopt the view that science is better understood through inquiry as recommended by the Singapore Ministry of Education (2008). During inquiry learning, students are actively engaged in science learning through activities like question-posing and answering, investigation, testing their hypothesis, evaluating and communicating their findings. Depending on the readiness of the students, the mode of inquiry can range from teacher centered (closed inquiry), teacher driven (very structured inquiry), teacher guided (guided inquiry) and student centered (open inquiry) which allows different strategies to be adopted.

Science simulations are probably suitable for these inquiry-based activities due to its capability of displaying multiple representations and its interactivity (C. H. Chen, Wu, & Jen, 2013). In using the simulations to conduct investigations, the students are learning together with technology (Jonassen, Howland, Marra, & Crismond, 2008). Using simulation as their partner in learning, these simulations can present the data in table or graphical form automatically. Hence, the students' cognitive resources are freed up so that they can now focus on higher-order thinking processes (e.g., analyzing the results and designing investigations). Furthermore, the simulation offers multiple representations (e.g., word, pictures, diagrams, graphs and table of values) of the same or related concepts. These multiple representations complement one another and students can thus now "integrate information from the various representations to achieve insights that would otherwise be difficult to achieve with only a single representation" (Wong, Sng, Ng, & Wee, 2011, p. 178).

With a deeper understanding of the concept, the students can now be in a better position to respond to questions, to evaluate and communicate their findings. In using the science simulations, students are allowed to manipulate the experimental variables to test their hypothesis. This allows for "genuine interactivity in terms of active learner contribution and engagement" (Hennessy, Deaney, & Ruthven, 2006, p. 702) . With this interactivity feature, students can now make use of the simulations and design their own investigation; by varying different parameters and collect their data accordingly to test their hypothesis.

With these affordances, students can now be actively engaged in inquiry-based activities (e.g., data collection and presentation, changing variables and hypothesis testing). Simulations are highly dynamic and have the potential to enhance students' learning gains (C. H. Chen & Howard, 2010; Lindgren & Schwartz, 2009; Scalise et al., 2011). By experimenting with the simulations, the students are learning by doing and are no longer just the passive receivers of information in a typical teacher-centered class setting. Such student-centered learning with technology is, of course, the much preferred approach for the contemporary researchers (Jonassen et al., 2008; S.-H. Liu, 2011).

**Reducing Cognitive Load in Science Simulations**

Cognitive load refers to the "strain that is put on working memory by the processing requirements of learning task" (Driscoll, 2005, p. 136). We believe well designed simulation is a key to supporting learning as students can focus on the learning task and not struggle with simulation. For example, too little information represented does little to aid learning (Blake & Scanlon, 2007; Y. L. Chen et al., 2011; van der Meij & de Jong, 2011) while too much information may result in too high a cognitive load to integrate information from the various representations (Sweller, 2005). Thus, in a well-designed simulation, the cognitive load should be minimized (Scalise, et al., 2011) as eliminating features "that are not necessary for learning will help students to focus on the learning processes that matter"(de Jong, 2010, p. 109). Here, we present reputed educational psychologist, Mayer (2008) 10 evidence-based principles to reduce cognitive load (See Table 1) .

*Table 1.* 10 evidence-based principles for reducing cognitive load Adapted from (Mayer, 2008)

| Purpose | Principle | Description |
|---|---|---|
| Reducing Extraneous Processing.<br><br>Extraneous Processing does not contribute to the achievement of instructional goal | Coherence | Remove extra information which is not needed achieve instructional goal |
| | Signaling | Highlight essential material |
| | Redundancy | Do not add text to narration |
| | Spatial contiguity | Place related representation (i.e., words, diagrams) close to one another<br>To prevent "split attention" effect |
| | Temporal contiguity | Present corresponding narration and animation concurrently |
| Managing Essential Processing<br><br>Essential Processing is required for the achievement of instructional goal. | Segmenting | Learner can control pace of animation |
| | Pretraining | Provide pretraining in the name, location, and characteristics of key components |
| | Modality | Present words as spoken text rather than printed text. |
| Fostering Generative Processing Generative processing involves organization and integrating material | Multimedia | Use words and pictures rather than words alone. |
| | Personalization | Use words in conversational style rather than formal style. |

Other than the principles suggested by Mayer (2008), we can also consider other ways to optimize working memory. According to a study by Ozcelik, Karakus, Kursun and Cagiltay (2009) , colour coding increased both retention and transfer performance. Such enhanced learning can be attributed to the ease of locating the related information (e.g., graphic or text) and hence leading to a better integration of the multimedia material.

**Research Gap in Elementary Science Simulation**

There is no doubt that both researchers and educators are interested in using simulations for classroom teaching. For example, high school students used electricity and collision cart simulations in the studies, of Liu & Su (2011) and Wee (2012) respectively, while researchers were investigating the effect of instructional support on learning with simulation for eighth graders (Eckhardt, Urhahne, Conrad, & Harms, 2013). However, there is little discussion about the use of simulations in elementary school settings. We conducted searches in the database Web of Science using descriptors "simulation” and "elementary school” and “science” for articles published since 2008 on April 19 2013. The search yielded only three results (Jaakkola & Nurmi, 2008; Jaakkola, Nurmi, & Veermans, 2011; Unlu & Dokme, 2011). Similar conclusion was also reported in the study of Smetana and Bell (2012) who examined 61 empirical studies in ERIC and Education Fulltext databases. Only two (3.3%) studies were conducted in elementary school setting. It is evident from these findings that there is indeed a paucity of research done in elementary school setting.

**Energy Simulation**

The conservations law of energy states that in any closed system (including the Universe), the total quantity of energy remains fixed shares the same big idea about fundamental nature such as conservation of momentum in a closed system, will always been both an universal and relatively difficult (Papadouris & Constantinou, 2011; Stylianidou, 2002) . This can be due to the abstract nature of the topic. We postulate that traditional pen paper teaching methods does not allow primary school students to apply this concept in other related areas like energy change in chemical reaction in their later schooling stage (Nordine, Krajcik, & Fortus, 2011). Therefore, we propose using well designed simulation with high interactivity and reduced cognitive load could afford for deeper learning as students can become “active agent in the process of knowledge acquisition” (Rutten et al., 2012, p. 137).

A search for energy-related simulations was conducted in the 5 popular simulations sites (NTNUJAVA Virtual Physics Laboratory, Java Applets on Physics , PhET Project, MyPhysicsLab and Interactive Physics and Math with Java) as suggested by Chen (2010) in her review on free simulation-based virtual laboratories. There are, of course, good energy simulations designed but they are not customized for use in elementary school (e.g., with extra information like acceleration due to free fall).

**Open Source Physics Simulation**

Open source physics simulations are characterized by access to the source codes. These codes are modified using the free Easy Java Simulation (EJS) available at http://www.um.es/fem/EjsWiki/Main/Download (see

). The move towards open source simulations  can be viewed as part of the increasingly popular movement  towards open content like the massive open online course (DeSilets, 2013) and open textbook (Wiley, Hilton, Ellington, & Hall, 2012). Open Content is one of the emerging technology trend highlighted in the reputed 2013 Horizon Report (Johnson et al., 2013). They described open content as "materials that are freely copyable, freely remixable, and free of barriers to access, sharing, and educational use" (p. 7) . This is very much similar to the spirit of open source physics in which the newly created or remixed simulations are shared with the community.

In this paper, we took an existing roller coaster simulation Open Source Roller Coaster (Gallis, 2010) and infused reducing cognitive loading principles using EJS.  In the spirit of open content, we released our primary school version here http://www.phy.ntnu.edu.tw/ntnujava/index.php?topic=2383.0.  Such enhancement to the existing energy simulation is, of course, much more efficient than creating a completely new simulation. Moreover, like the other open source physics simulations, the energy simulation is built by physics professors modeled using Physics equations. These simulations are more realistic thus aid the understanding of scientific concepts better Blake and Scanlon (Blake & Scanlon, 2007).

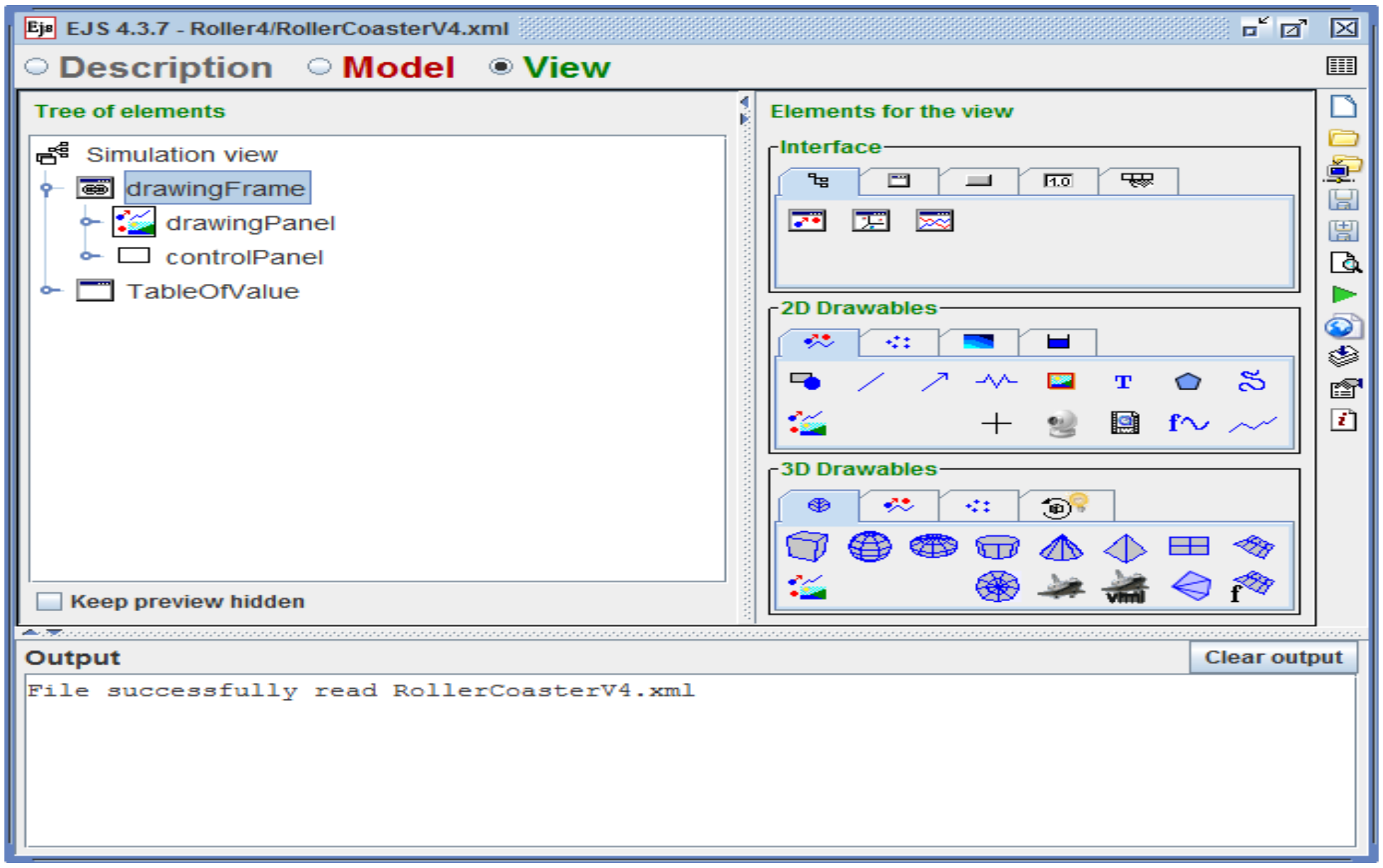

*Figure 1.* Easy Java Simulation (EJS) Toolkit

## TPACK and Open Source Energy Simulation

Koehler and Mishra (2005) first mooted technological pedagogical content knowledge (TPACK) which extends Shulman's conceptual model (1987) of Pedagogy Content Knowledge (PCK). TPACK refers to "refers to the synthesized form of knowledge for the purpose of integrating ICT/educational technology into classroom teaching and learning" (Chai, Koh, & Tsai, 2013, p. 32). As a framework, the constructs of TPACK are technological knowledge (TK), pedagogy knowledge (PK), and content knowledge (CK), Pedagogy Content Knowledge (PCK), Technology Content Knowledge (TCK), Technology Pedagogy Content (TPK) and Technology Pedagogy Content Knowledge (TPACK). The TPACK framework can be used in the design of teacher education program  for designing technology-based lessons (Chai et al., 2013)

One possible strategy to develop teachers' capability in infusing technology in classroom teaching is through the design of the educational technology to be used in authentic situations (Chai, Koh, Tsai, & Tan, 2011; Koehler & Mishra, 2005; Voogt, Fisser, Pareja Roblin, Tondeur, & van Braak, 2013). In a similar manner, designing open source simulation allows the teachers to experience the rich interplay of the relationships between technology, pedagogy and content (Koehler, Mishra, & Yahya, 2007). In the use of open source simulations, they can even have a richer experience as they are now involved in modifying the simulation to suit their needs. In such modification, teachers can deepen their TPACK as the design can be seen as an "a process of weaving together components of technology, content, and pedagogy" (Koehler & Mishra, 2005, p. 135). They need to integrate the technology knowledge (what EJS can do) and pedagogical content knowledge (i.e., how to teach energy concept). TPACK, as knowledge, in this case, can transform teaching of science as the teacher is now able to customize the simulations to meet their specific pedagogical needs. Such customization by teachers is rarely published as the codes are not easily available or there is not relatively simple to use simulation author toolkits (like EJS or Molecular Workbench).

In remixing the simulation, the teachers are taking on the dual role of both the teacher and the designer. The dual role is advantageous, as the intention of the designer and the teacher will definitely be aligned. Teachers know the learning needs of the students in class and are clearly in the best position to design the resources to best fit their classroom practice (Hjalmarson & Diefes-Dux, 2008). Such tailored-made simulation is aligned to the students' need and intended learning outcomes and thus are more likely to be adopted by teachers (Moizer, Lean, Towler, & Abbey, 2009). As the teachers are involved in designing simulation, they were, of course, aware of the affordances of the simulations and thus were able to better utilize them in the class. With holistic view of the relationship and the overlapping of the technology, pedagogy and content knowledge, they were now in a better position to use the simulation meaningfully in class.

## Stages in Using Open Source Energy Simulation

In this study, we proposed classifying the use of open source simulation into 3 stages: design, customization and implementation. Such classification can possibly guide the interested educators and researchers in the use of open source simulation. In the design stage, we decided on the possible features to be added. This was a paper design exercise before the simulation was actually changed using on EJS. Based on features, the simulation that was close to what was required was chosen. This would ensure that the open source simulation could be customized in the shortest possible time. In this case, the Open Source Roller Coaster (Gallis, 2010) was selected. Moreover, students were more likely to connect the simulation experience to the roller coaster real life experience.

The customized simulation was implemented in an enrichment lesson. 35 Grade 4 and Grade 5 mixed-ability students took part in this study. With no prior knowledge on energy, the lesson served as an introduction to energy. The objective of the lesson was to investigate energy conversion (in this case, it is the conversion of Gravitational Potential Energy to Kinetic Energy and vice versa). Energy simulation was used to support inquiry-based activities (i.e. like investigate the relationship between height and potential energy). The simulation can be downloaded at : https://dl.dropbox.com/u/24511248/JaveEJS/ejs_RollerCoasterV4.jar.
A guided inquiry-based lesson with the instructional guidance (i.e., reinforcing the energy concept and introducing simulation) was adopted rather than the pure discovery approach (Eckhardt et al., 2013). Learning is likely to be best supported by "instructional guidance rather than pure discovery, and curricular focus rather than unstructured exploration" (Mayer, 2004, p. 14). Similarly, in the study of simulations on middle school students, C. H. Chen and Howard (2010) also suggested that appropriate instructional guidance could enhance students' learning experience. This study was conducted over two 2.5-hr lessons using the popular constructivist-based BCSE 5E Model (Bybee et al., 2006). See Table 4 for details.

*Table 2.* Lesson Implementation using BCSE 5E Model

| Phase | Description | Activities Implemented |
|---|---|---|
| Engage | Elicit their prior understanding | You Tube Videos |
| Explore | Develop their understanding | Exploring the Energy Simulation |
| Explain | Demonstrate their understanding | Google Form Submission and Inquiry-based Worksheet |
| Elaborate | Deepen their understanding | Roller Coaster (Figure 9) to challenge the students' perceived perception on energy |
| Evaluate | Assess their understanding | Change the settings of the Roller Coaster Simulation or Pendulum Simulation (**Error! Reference source** |
| | | **not found.**) to record the highest kinetic energy |

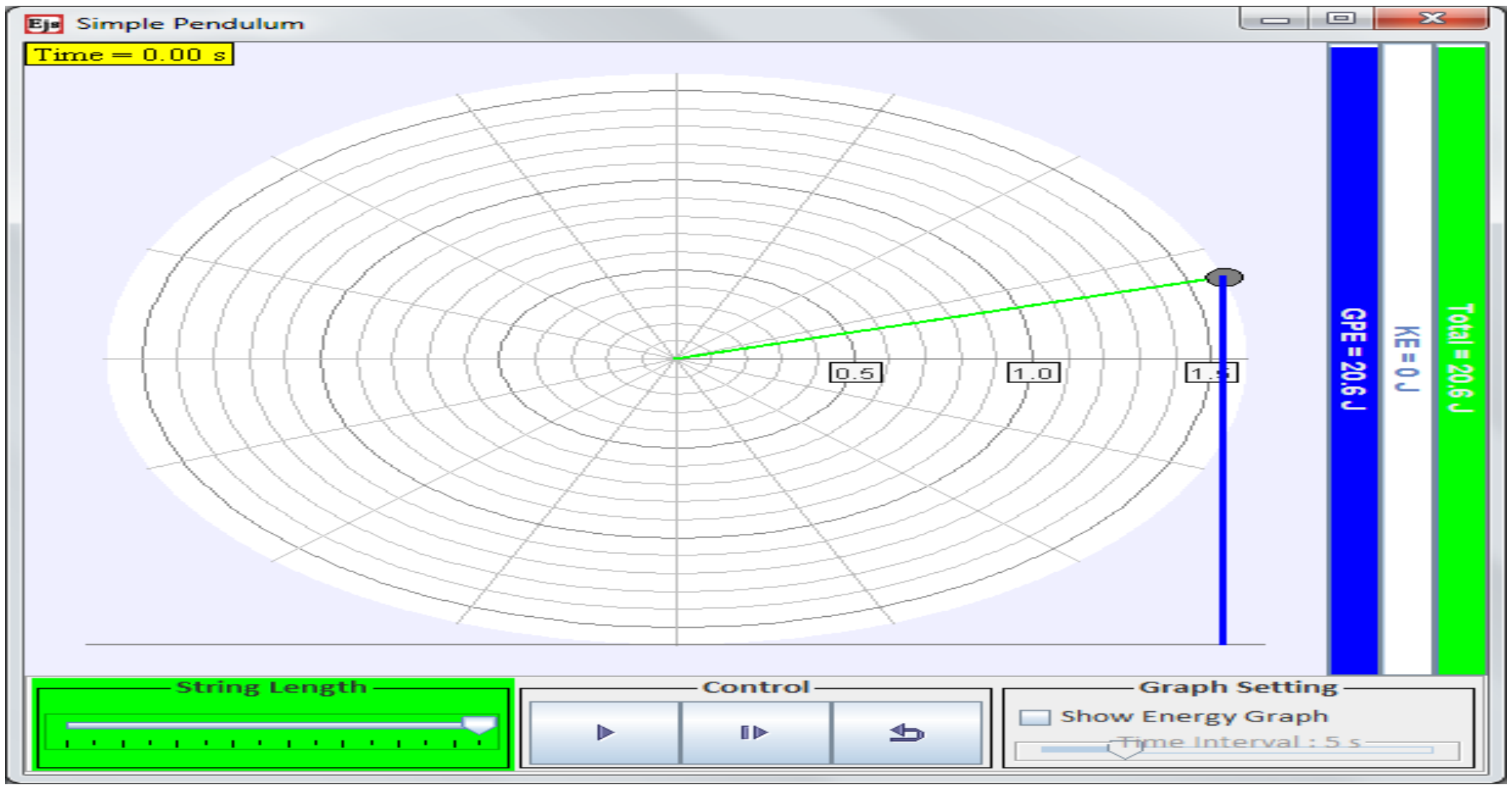

*Figure 2.* Pendulum Simulation

## Purpose of the study

The purpose of this exploratory pilot study was to document how we designed, customized and implemented the simulation in an elementary class. It sought to provide evidences of increased interactive engagement the use of simulation both from the teachers and the students' perspectives. This study would be of interest to both teachers and researchers who want to adapt and use simulations or teach others how to modify the simulation. The study was guided by the following questions:

1. What are the generic principles and its corresponding TPACK construct that guides the design and customization of simulation?
2. What are the students' learning experiences with simulation?

## Research Design

A case study approach was adopted so as to seek a deeper understanding on how the open source energy simulation was adapted and used in class. In this study, the case represented the stages of design, customization and implementation of the energy simulation in an elementary school in Singapore.

To develop converging lines of inquiry, multiple sources of evidence were collected (Yin, 2008). Furthermore, findings derived and triangulated from the different data source are likely to be deemed more reliable and convincing. More details on data source and analysis and the corresponding process can be found in *Table 3*. Both directed content and conventional content analysis were used for qualitative data (Hsieh & Shannon, 2005). For example, predetermined categories derived from the literature review on principles to reduce cognitive load served as a guide in analyzing the simulations during the customization stage.

*Table 3.* Data Source and Analysis

| Stage | Data Source | Data Analysis |
|---|---|---|
| Design | Design Notes<br>Energy Simulation<br>Other Simulations<br>Other research articles | Conventional content analysis (emerging categories) |
| Customization | Energy Simulation<br>Other Simulations | Directed content analysis (predetermined and emerging categories) |
| Implementation | Field notes on lesson observations<br>Students' Google form submission (Open ended question)<br>Focus Group Discussion with 12 students (from different abilities)<br>Students Perception survey (Comments) | Conventional content analysis |
| | Students Perception survey (5-point Likert Item)<br>Students' Worksheet | Descriptive Statistics |

## Results and Discussion

### PK in Design Stage

The design features added in the simulation was noted (see *Table 4* and *Figure 3*). They were further studied to surface the generic underlying principles and distill to the generic TPACK construct that could help guide the design process of other similar simulation. In this case, it is the pedagogy knowledge (PK) which is the knowledge "to teach a subject matter without references towards content" (Chai et al., 2013, p. 33).

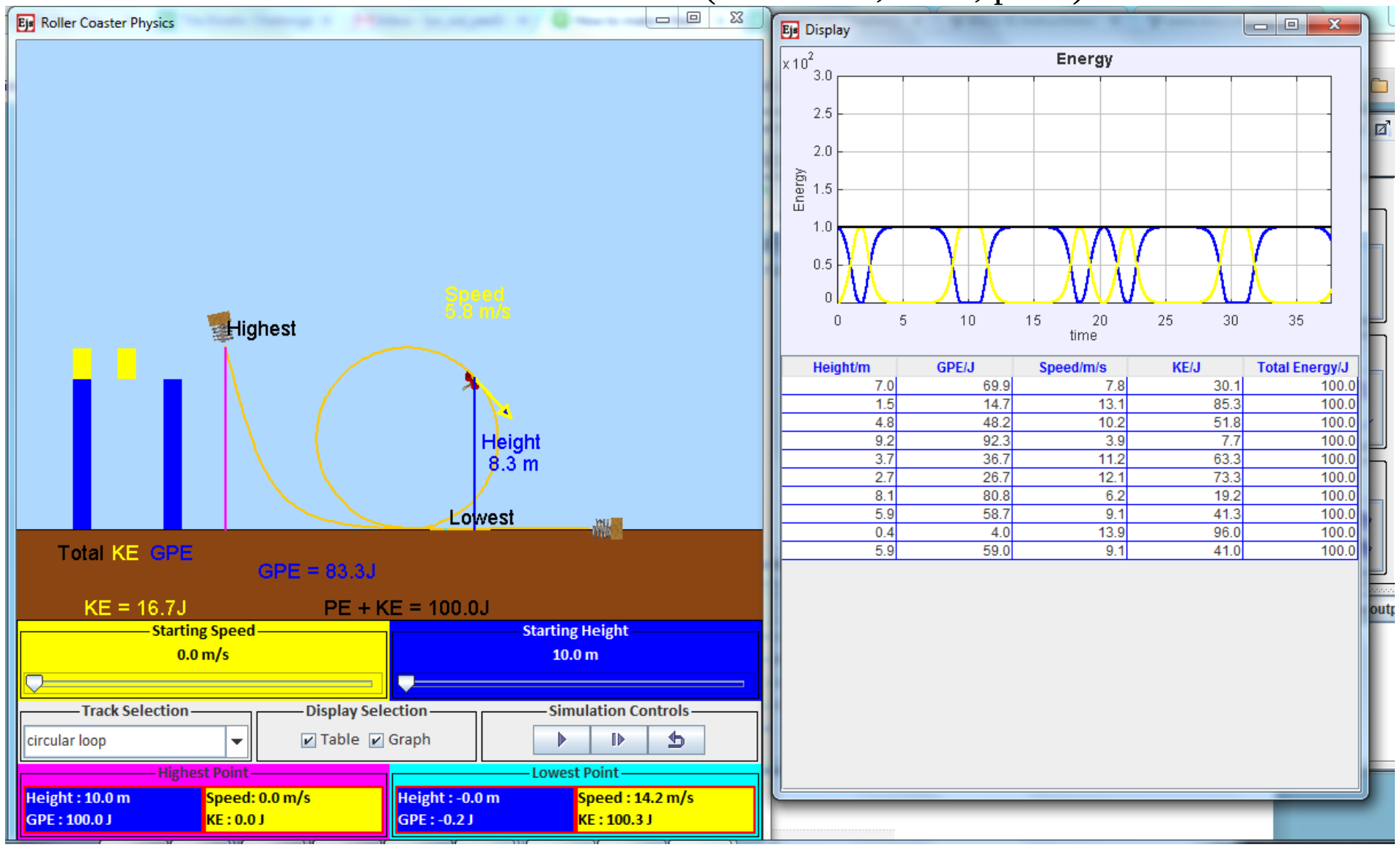

*Figure 3*. Modified Roller Coaster

*Table 4*. Design

| Features | Principles | TPACK Construct |
|---|---|---|
| Information Box about highest point and lowest point | Learning outcome | PK |
| Relationship between height and gravitational potential energy, speed and kinetic energy | Learning outcome | PK |
| Display of Graph and Values | Inquiry-based activities | PK |
| Controls for variables (like starting height and the mass) | Inquiry-based activities | PK |
| Added in Circular Loop | Teachers' feedback | PK |
| Display of maximum and minimum height on the graphic | Teachers' feedback | PK |
| Display of real GPE and KE | Fidelity towards actual phenomenon: | PK |
| Newly designed roller coaster | Students' misconception | PK |

From the analysis, it was evident that pedagogical knowledge guided the design as it was our intention to make the simulations educationally sound. Our other simulations, design notes were also examined for the principles emerging from the analysis of the current simulation. This helped to triangulate the findings. The design principles were described below:

1. Learning outcomes: The simulation was modified such that the learning outcomes (like information about highest point and lowest point) can be made more explicit from the simulation. It will aim to persuade and convince the students of the intended concept based on their observation. This is hardly surprising as this is the primary purpose of the customization is to suit the localized setting (a Singapore elementary school). This factor was evident in the other collision cart (Wee, 2012) in which the momentum equation was added to the existing simulation (*Figure 4* )
2. Inquiry based activities: The simulation was further modified so that it could support inquiry-based activities (e.g., investigation, testing their hypothesis and evaluation of their findings). This was achieved by providing graph display and table of values, or given the students to control the variables like starting height. Similarly in the friction simulation, students could control the mass and the applied force (Figure 5) while values were displayed in collision cart simulation (*Figure 4*).

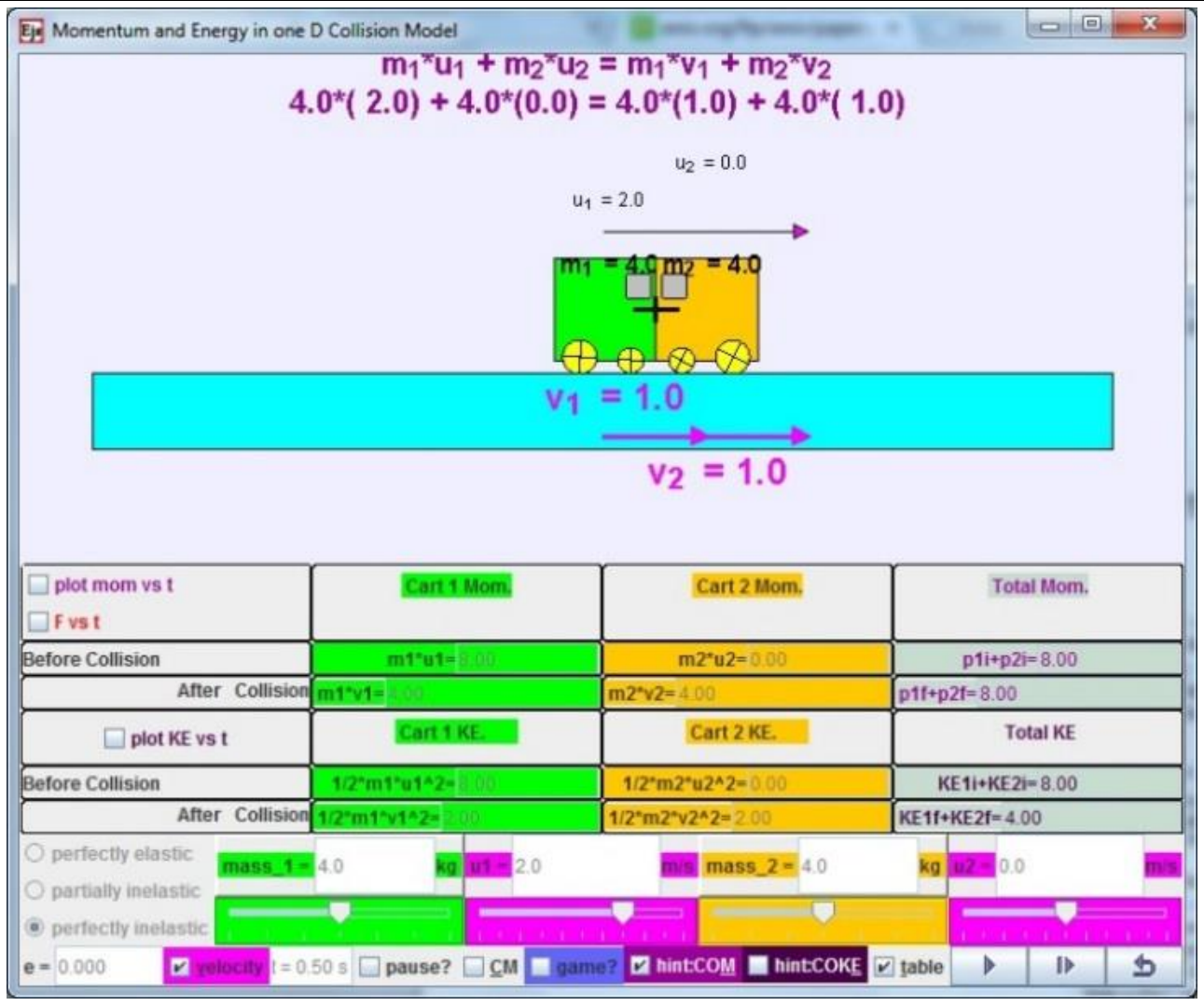

*Figure 4.* Collision Cart

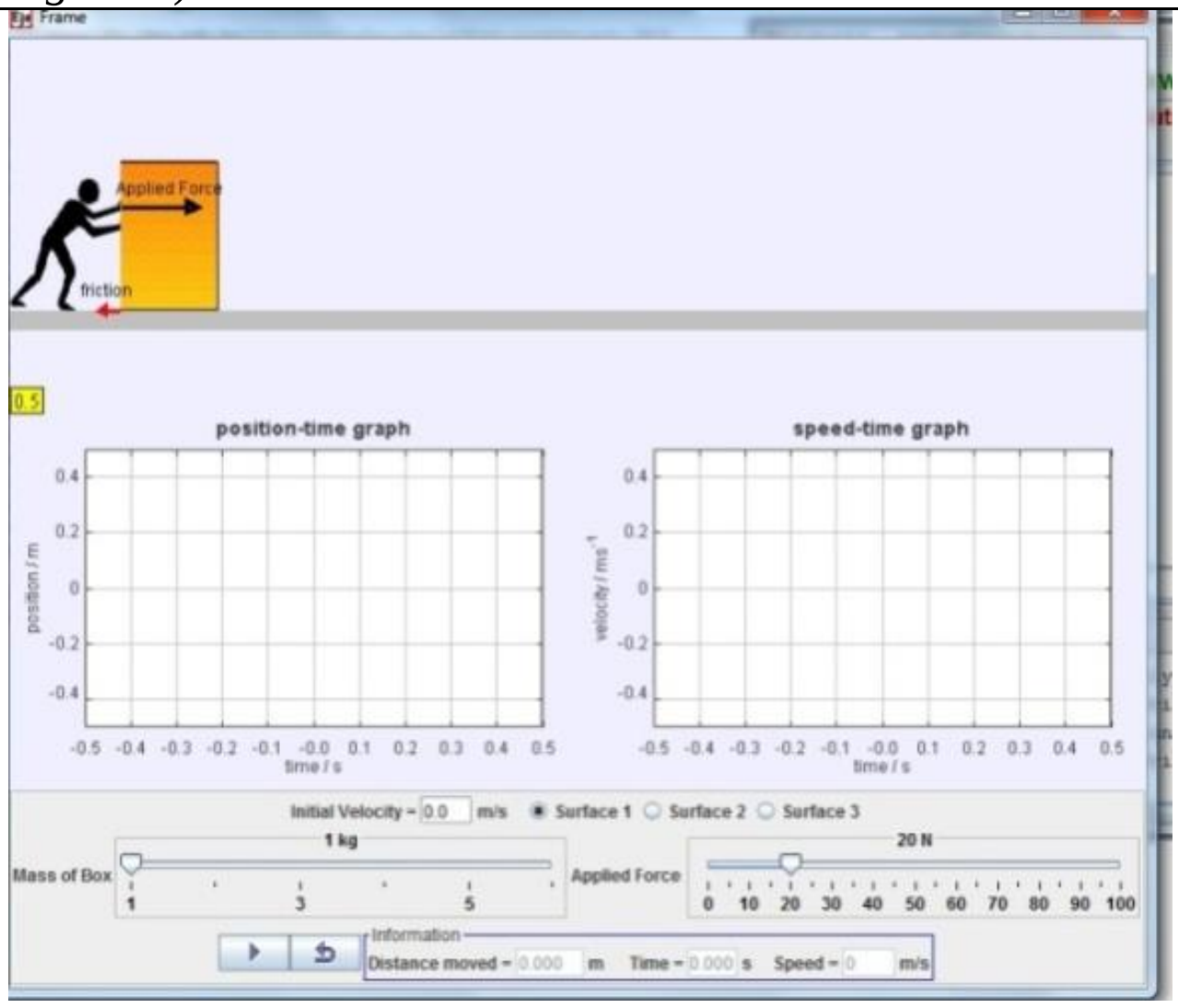

Figure 5**.** Friction Simulation

3. Feedback from other teachers: Feedback was solicited from other teachers regarding our first version of the design via an online survey. We acted upon the feedback and further enhanced the simulation. Such practice was adopted in other simulations too, as reflected by the designer notes for the Ripple Tank Simulation and Cooling/Heating Simulation (Figure 6). Such feedback was based on pedagogical reasoning so that students were able to learn better with simulations.

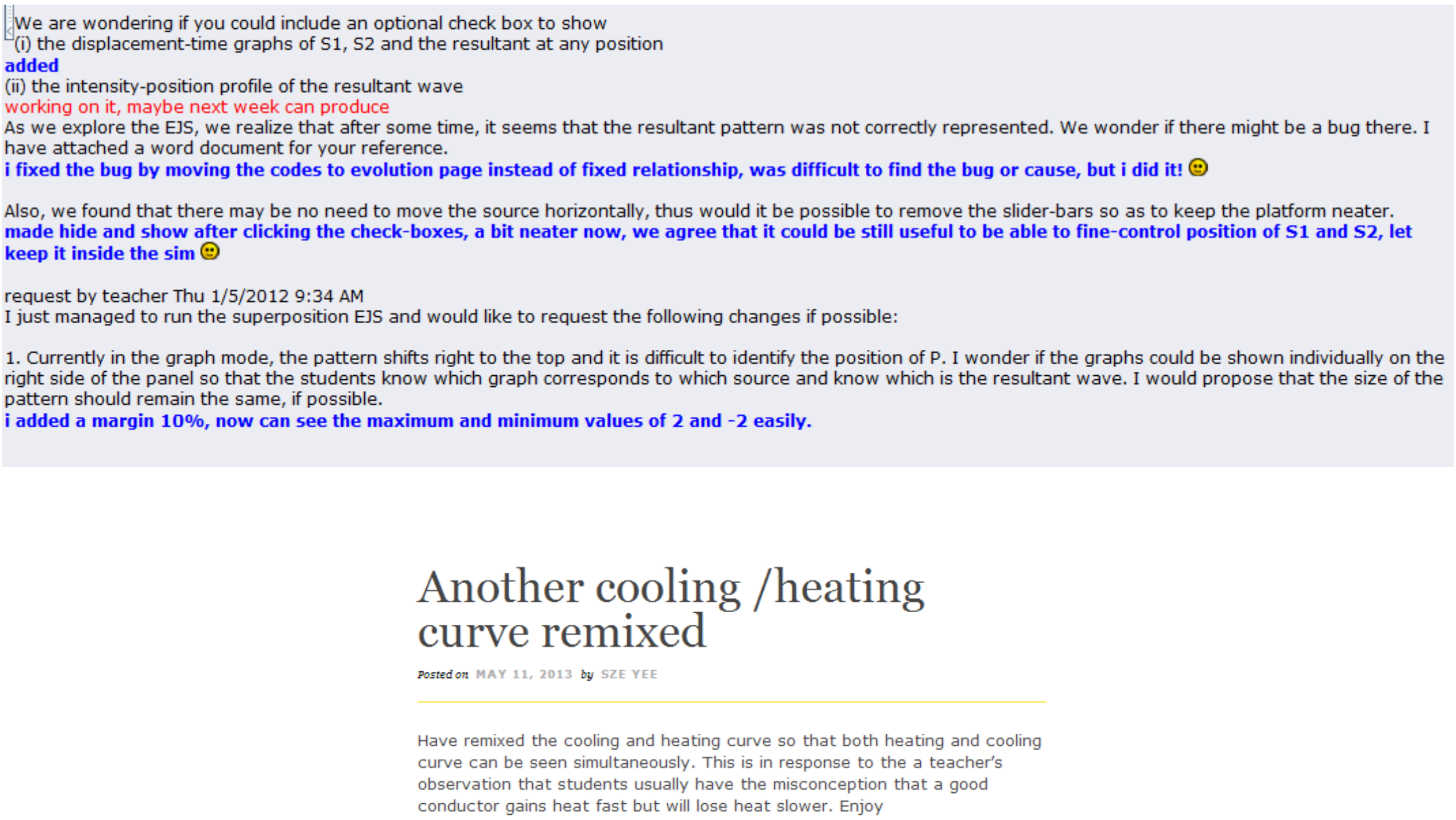
We are wondering if you could include an optional check box to show
(i) the displacement-time graphs of S1, S2 and the resultant at any position
**added**
(ii) the intensity-position profile of the resultant wave
working on it, maybe next week can produce
As we explore the EJS, we realize that after some time, it seems that the resultant pattern was not correctly represented. We wonder if there might be a bug there. I have attached a word document for your reference.
**i fixed the bug by moving the codes to evolution page instead of fixed relationship, was difficult to find the bug or cause, but i did it!** ☺

Also, we found that there may be no need to move the source horizontally, thus would it be possible to remove the slider-bars so as to keep the platform neater.
**made hide and show after clicking the check-boxes, a bit neater now, we agree that it could be still useful to be able to fine-control position of S1 and S2, let keep it inside the sim** ☺

request by teacher Thu 1/5/2012 9:34 AM
I just managed to run the superposition EJS and would like to request the following changes if possible:

1. Currently in the graph mode, the pattern shifts right to the top and it is difficult to identify the position of P. I wonder if the graphs could be shown individually on the right side of the panel so that the students know which graph corresponds to which source and know which is the resultant wave. I would propose that the size of the pattern should remain the same, if possible.
**i added a margin 10%, now can see the maximum and minimum values of 2 and -2 easily.**

## Another cooling /heating curve remixed

Posted on MAY 11, 2013 by SZE YEE

Have remixed the cooling and heating curve so that both heating and cooling curve can be seen simultaneously. This is in response to the a teacher's observation that students usually have the misconception that a good conductor gains heat fast but will lose heat slower. Enjoy

*Figure 6*. Teacher's feedback

4. Fidelity towards actual phenomenon: The simulations must be of correct level of representation (not too complex or simplified for the targeted audience) to bring about a meaningful and deep learning experience for the students (Blake & Scanlon, 2007). Open Source Physics simulations are modelled closely after the real-life phenomenon as they were based on Physics concepts. Similarly, in our modification of the energy simulations, we would base our codes on the Physics concepts. The gravitational potential energy (GPE) changed accordingly to height according the GPE equation (Figure 7). For the collision cart simulation, Newton's 3rd Law was clearly evident in the force graph (*Figure 8*).

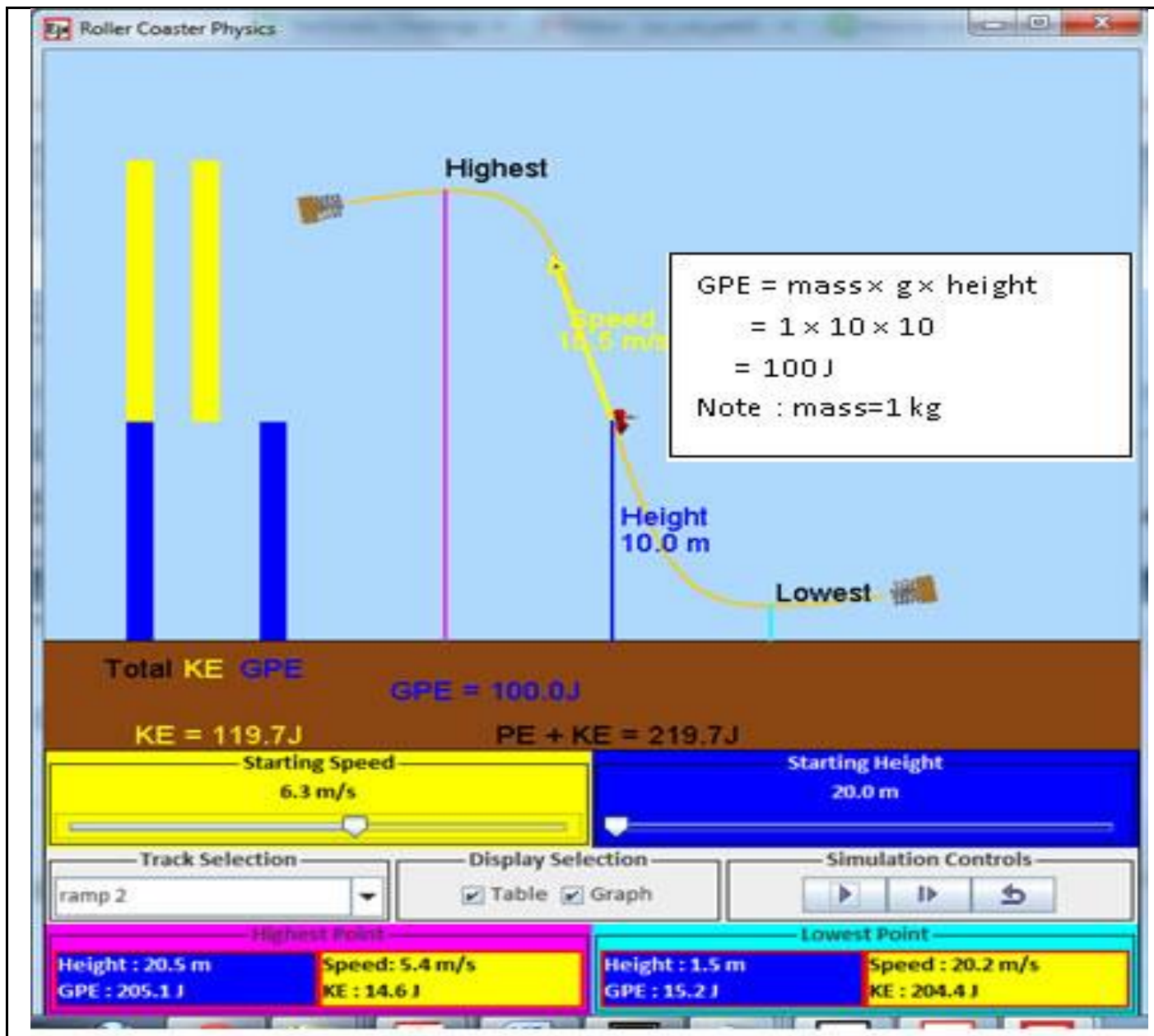

*Figure 7* . Energy Law in Roller Coaster Simulation

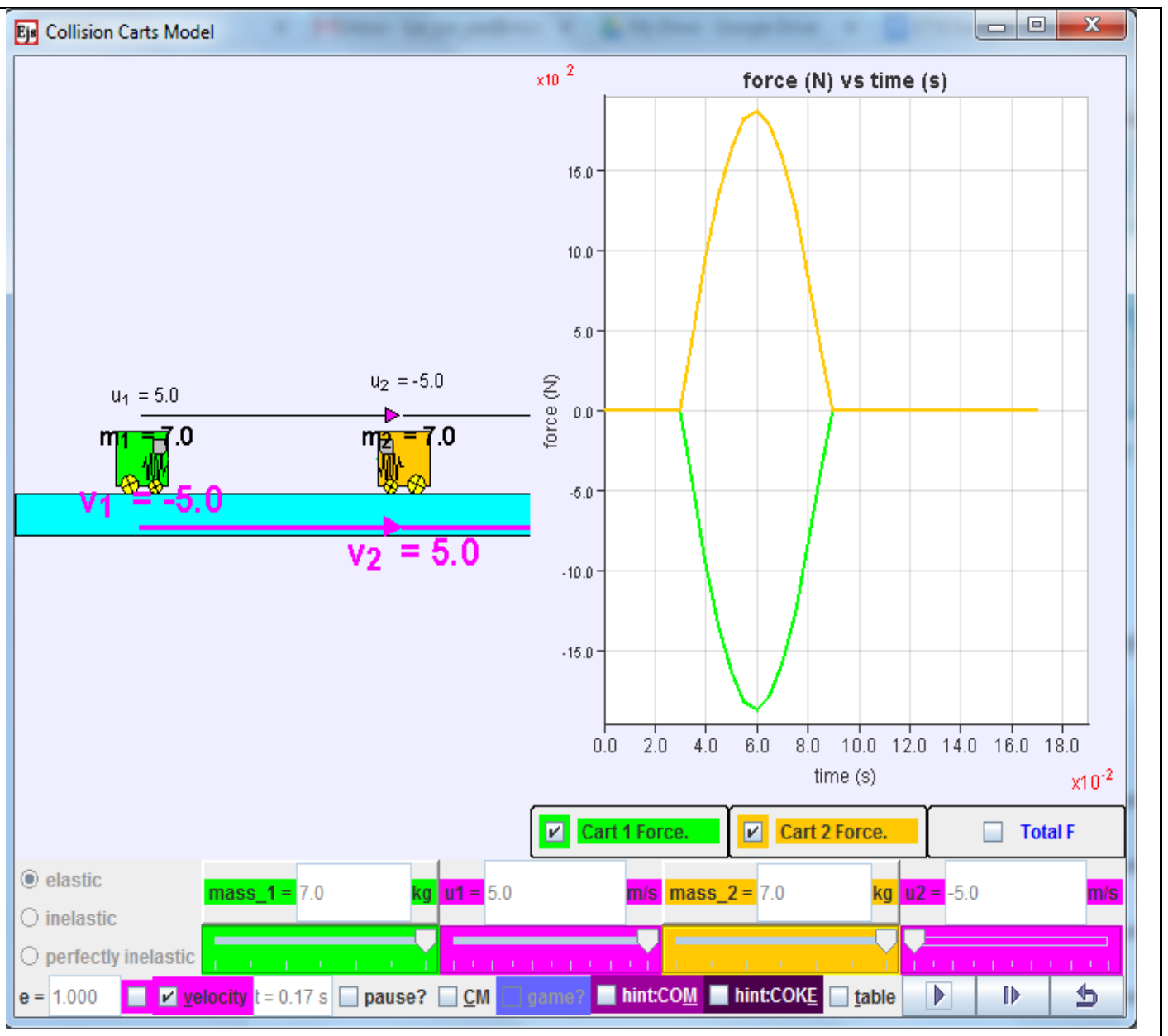

*Figure 8*. Newton Third Law in Collision Cart

5. Students' misconceptions: Based on our collective teaching experience, we surfaced the possible misconceptions that students were likely to have. They would usually assume that the object can only reach the same height as the initial position. Another roller coaster (with the initial position lower than the highest position) was created to challenge the students. They were to make the roller coaster reach the highest point. This allowed the students to experience cognitive dissonance so as to bring about a conceptual change (Y. L. Chen, Pan, Sung, & Chang, 2013; Smetana & Bell, 2012).

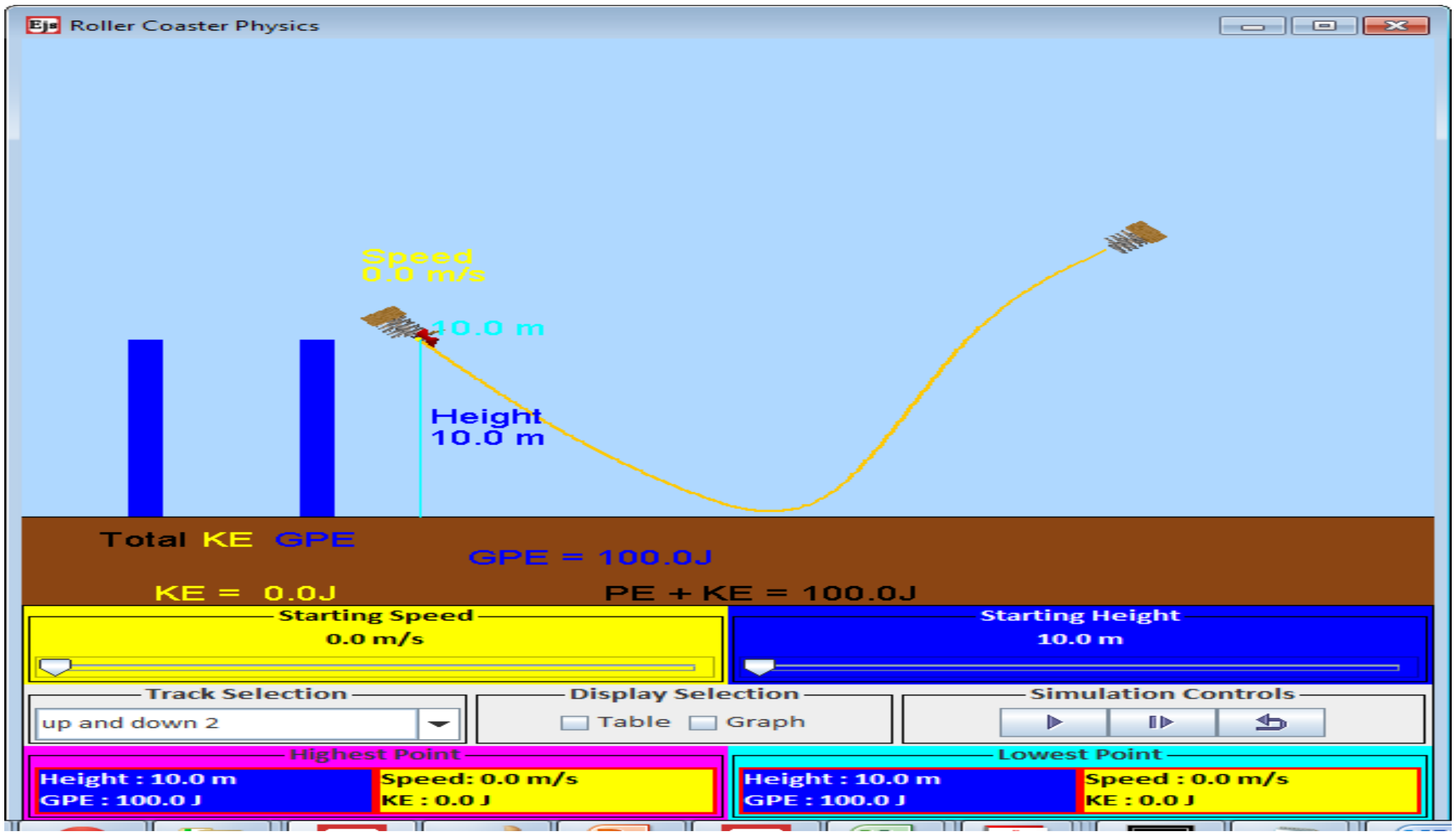

*Figure 9* . New Roller Coaster Design

**TPK in Customization Stage**

**The energy simulation was examined using the principles for reducing cognitive load (see** Reducing Cognitive Load in Science Simulations). Some principles are not applicable in the design of the simulation (i.e., narration-related principles). Clearly, such principles belong to the technology pedagogy knowledge (TPK) as we used the technology (what EJS can do) to make the simulation more pedagogically sound. TPK is the "knowledge of the existence and specifications of various technologies to enable teaching approaches without reference towards subject matter"(Chai et al., 2013, p. 33) and hence the similar principles can be adopted for customizing other simulations. To triangulate the findings, other simulations we customized were also inspected using the same lens. Please refer to *Table 5* for more details.

*Table 5*. Customization

| Customization Principle | TPACK Construct | Energy Simulation | Other Simulation |
|---|---|---|---|
| Coherence | TPK | Acceleration due to gravity (g), normal reaction force (R), energy loss (k) are removed. Information does not contribute to the learning outcome (<br>) | Forces (e.g., normal reaction force) removed in the inclined ramp simulation for elementary school (Figure 10. Roller Coaster Simulation<br><br>) |
| Signaling | TPK | Indicate the highest and lowest height in the graphic display (<br>). | Indicate the height of the ramp in the inclined ramp simulation(Figure 10. Roller Coaster Simulation<br><br>) |
| Spatial contiguity | TPK | Indicate the highest and lowest point in the graphic display<br>Display of velocity and height information in symbolic (length of arrow) and numerical form near the object.<br>Modification of Total Energy bar such it is made up of KE and PE energy bars<br>Change the orientation of KE bar such that it is moving downwards (<br>). | Display of velocity information in symbolic (length of arrow) and numerical form near the object in the inclined ramp simulation. The length of the arrow depends on the velocity<br>(Figure 10. Roller Coaster Simulation<br><br>) |
| Colour Coding | TPK | Consistent colour scheme for concepts related to potential energy and kinetic energy. | Consistent colour scheme in collision cart simulation (<br>Figure *11*. Inclined Plane Simulation<br>) |

Original Simulation Customized Simulation

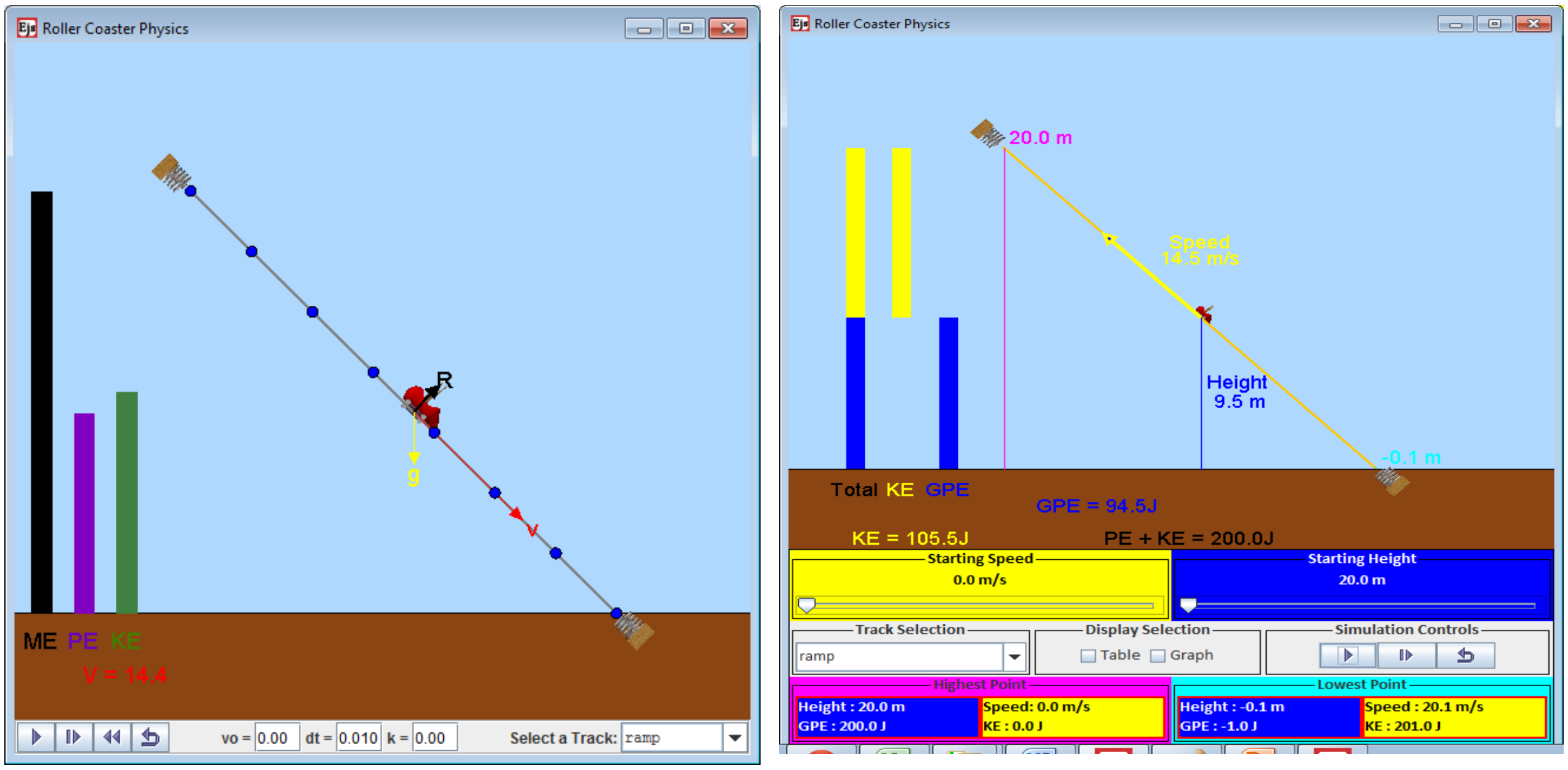

*Figure 10*. Roller Coaster Simulation

Original Simulation Customized Simulation

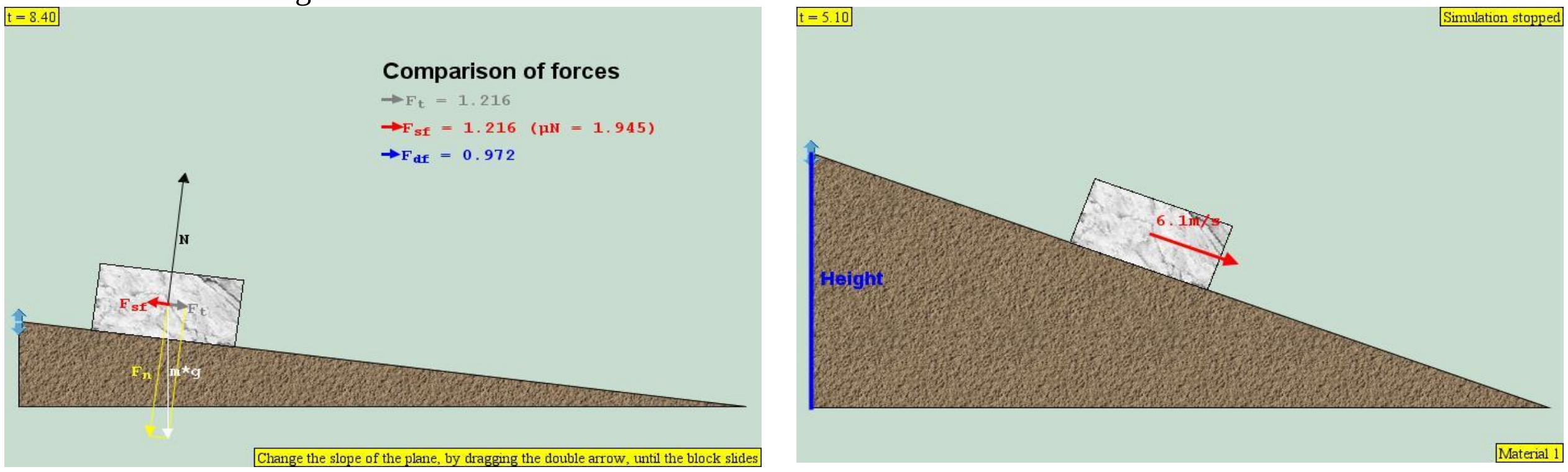

*Figure 11*. Inclined Plane Simulation

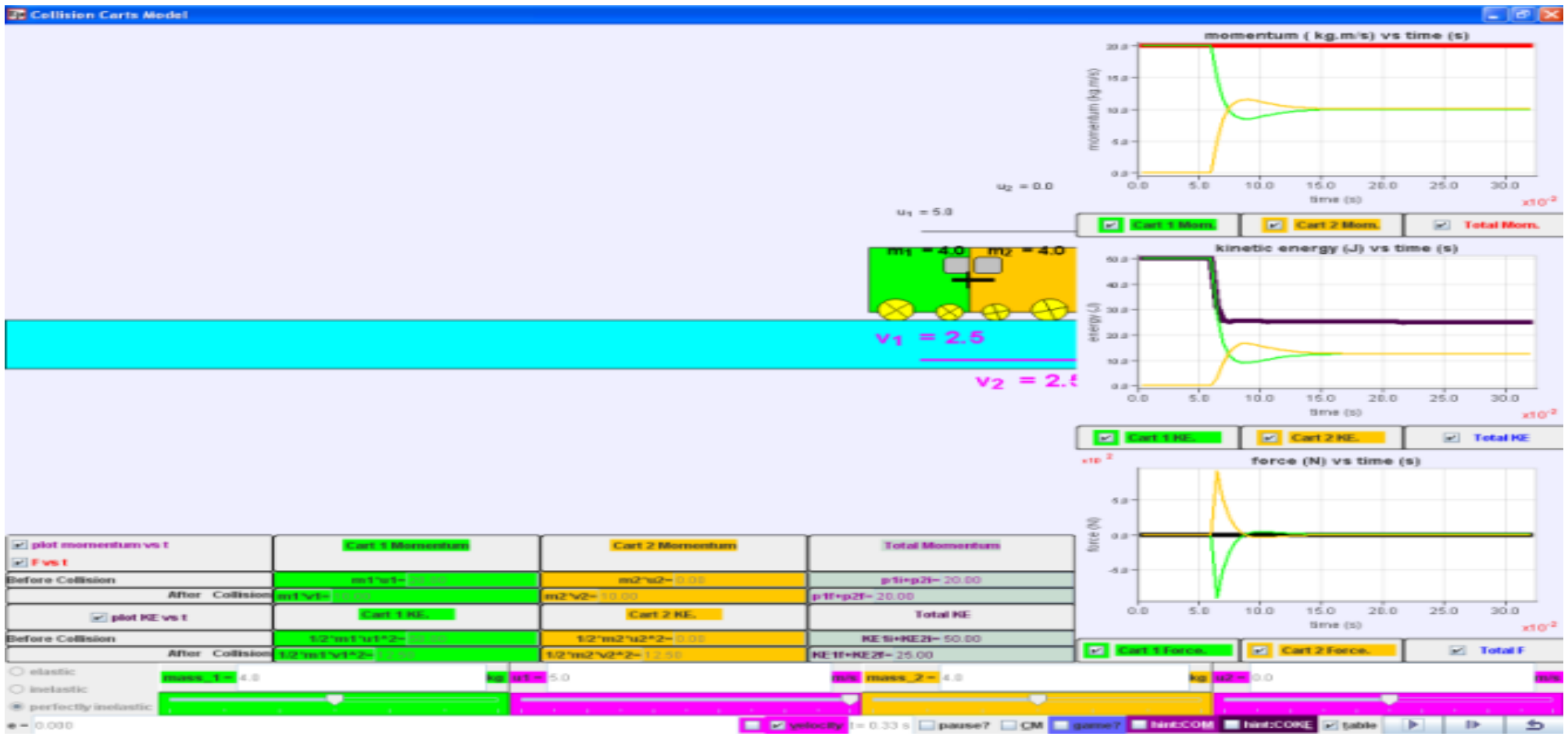

*Figure 12*. Collision Cart

## Learning Experience in Implementation Stage

The qualitative data (i.e., field notes, focus group discussion and the Google form submission) were examined to uncover the possible emerging categories for students' learning experience. From the analysis, the students had a positive learning experience as they found the features of simulation useful and were engaged in self-directed learning. Such conclusion is line with literature review on simulation as described in the earlier section on Inquiry Learning with Science Simulation. The findings were further supported by the survey which was on 5 – point Likert Scale and the students' worksheet. Out of the possible 35 students, 27 students completed the survey while 35 students' worksheets were collected.

## Positive Learning Experience

From field observations, comments in the survey and focus group discussion, students were engaged in learning with the simulations. They had no trouble navigating and exploring the simulation. Most of them found the lesson "cool" and "fun" and used the term "play" to describe their learning experience.

As the roller coaster could be customized, they were investigating the energy relationship using different way. Most students were able to articulate the energy principles during field observation. This observation also concurred with the Google Form Submission. Students also showed their understanding of the concepts in the hard copy worksheets. Majority of the students (30) were able to state the energy conversion correctly.This positive learning experience also concurred with the perception survey in which the students enjoyed and found it easy to use and learned with the simulation (see *Table 6*).

*Table 6*. Positive Learning Experience

| | Mean | SD |
|---|---|---|
| I enjoy learning about science using simulation. | 4.15 | 1.06 |
| I am able to learn about energy on my own using the simulation. | 3.74 | 1.11 |
| I find it easy to use the simulation. | 3.89 | 1.19 |
| Reliability Test: Cronbach's Alpha | 0.886 | |

**Features**

The features (e.g., colour scheme, bar graph, numerical information) conceived during the design and the customization stages helped the student in understanding the energy concepts (see *Table 7*). The survey findings were further supported by focus group discussion. The representative comments were:

> *The indication of height (blue line) helps me to see the relationship between GPE and height*
>
> *KE depends on the speed. The yellow arrow on the simulation shows that when the speed is low (like 7.3 m/s), the KE is 26.3J. But when the arrow is longer (faster) (like 11.3m/s) the KE is 60.5J.*
>
> *The total energy is made up of KE and GPE. The bar at the sides shows the bar. The TE bar shows both KE and GPE*

However, the weaker student during focus group discussion indicated that they would need teachers' guidance to point out the design features (e.g., consistent colour scheme) so that they could discover the energy relationship on their own.

*Table 7.* Design Features

| | Mean | SD |
|---|---|---|
| The colour scheme helps me to understand the energy better. | 4.41 | 0.97 |
| The bar graphs helps me to understand the relationship between KE, GPE and Total energy. | 4.33 | 1.04 |
| The numerical information helps me to understand the relationship between KE and GPE | 3.96 | 1.12 |
| The indication of height (blue line) helps me to see the relationship between GPE and height. | 4.18 | 1.08 |
| The indication of speed (yellow line) helps me to see the relationship between KE and speed. | 4.30 | 1.03 |
| Reliability Test: Cronbach's Alpha | 0.932 | |

**Self-Directed Learning with Simulation**

Analysis from the various data sources indicated that students were engaged in self-directed learning with simulation. They enjoyed the interactivity and the experimentation as one student commented in the survey "*I like the fact that we can experiment with the simulation by ourselves, and the simulation is very interactive."* The focus group discussion also revealed similar results with students enjoying the process of controlling the variables and evaluating their hypothesis on their own. Even if they did not fully comprehend what the teacher had explained, they could make use of the simulation to monitor their learning as one student pointed out "*We don't understand we can check ourselves*".

It was interesting to note that students were evaluating their hypothesis in way we had not thought of. For example, some students used the starting speed and starting height to discover about the kinetic and gravitational potential energy:

> *If I adjust the height where the roller-coaster starts from, the amount of gravitational potential energy would be higher/lower as shown on the bar.*

From field observations, the students figured out how to customize the roller coaster on their own. They enjoyed the customization process as evident from the perception survey (M=4.24, SD=1.23). They even began experimenting with different variables (i.e. starting speed and starting height) so that the roller coaster could arrive at a higher point than the initial position. Most students were able to address their own alternative misconception before they were addressed formally. From Google Form submission, the students showed that they understood that for the object to have maximum kinetic energy; they had to "*make its starting height very high*". Their experience was aptly put across by one student during the focus group discussion:

> *I enjoy customizing the stimulation as I change the ramps, height and speed according to my own fancy and changing the ramps helps me know how I can get from the start point to the end point quickly/slowly, and figure out why some ramps don't allow me to arrive at the end point as quickly as the other ramps.*

## Future Work

Simulations can be used to provide differentiated instructions. With EJS, different versions of the same simulation can be created to cater to the different needs of the students. The expertise reversal effect can be considered to guide the design of the different versions of the simulation. In expertise reversal effect, students with different prior knowledge need different design features as features that "are effective with low-knowledge individuals can lose their effectiveness and even have negative consequences for more proficient learners"(Kalyuga, 2007, p. 510). During the focus group discussion, the weaker students found the graph difficult to understand as they "*don't understand what the graph is moving for*". They would prefer the table of values (Figure 13).

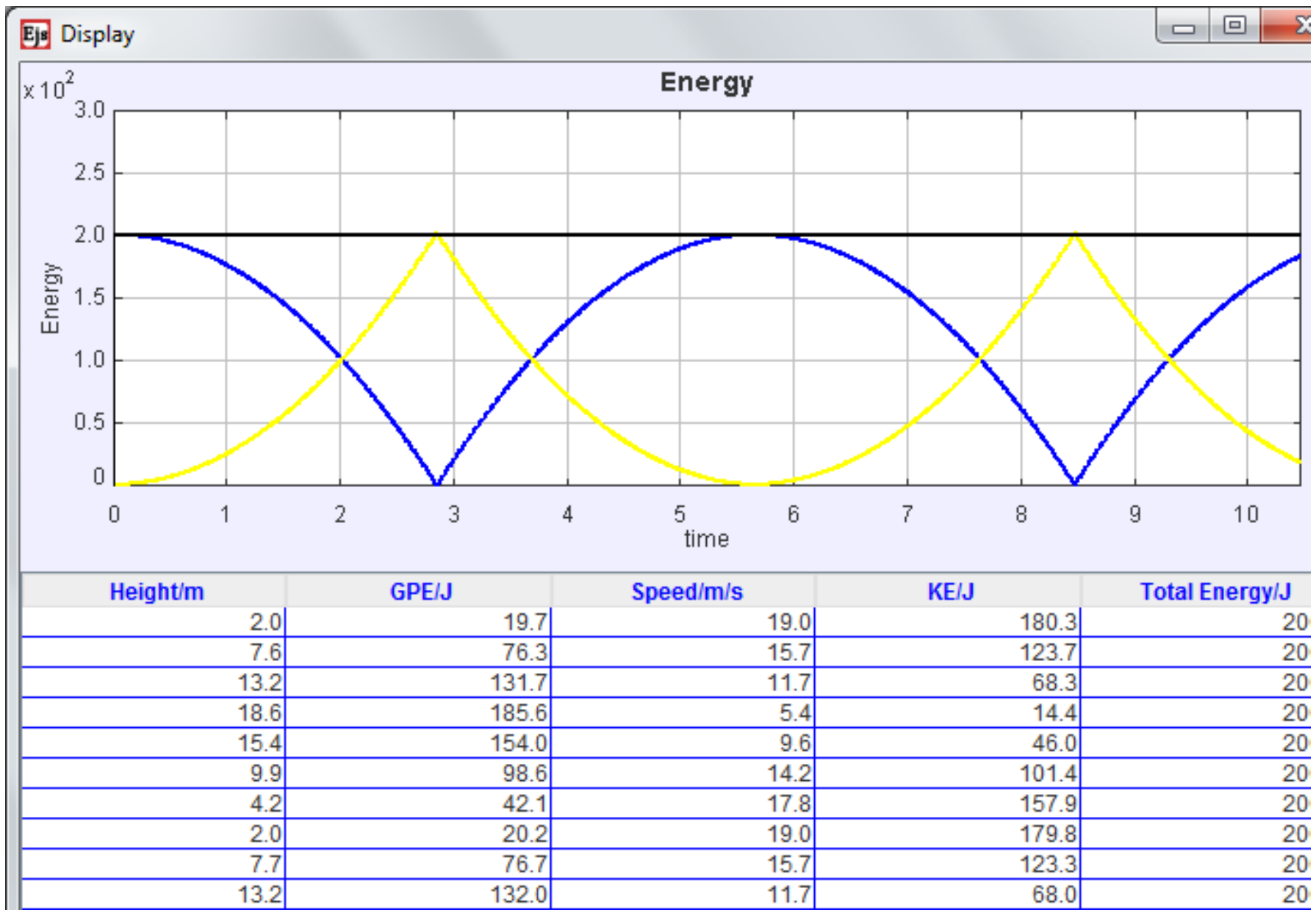

| Height/m | GPE/J | Speed/m/s | KE/J | Total Energy/J |
|---|---|---|---|---|
| 2.0 | 19.7 | 19.0 | 180.3 | 20 |
| 7.6 | 76.3 | 15.7 | 123.7 | 20 |
| 13.2 | 131.7 | 11.7 | 68.3 | 20 |
| 18.6 | 185.6 | 5.4 | 14.4 | 20 |
| 15.4 | 154.0 | 9.6 | 46.0 | 20 |
| 9.9 | 98.6 | 14.2 | 101.4 | 20 |
| 4.2 | 42.1 | 17.8 | 157.9 | 20 |
| 2.0 | 20.2 | 19.0 | 179.8 | 20 |
| 7.7 | 76.7 | 15.7 | 123.3 | 20 |
| 13.2 | 132.0 | 11.7 | 68.0 | 20 |

*Figure 13.* Graph and Table of Value

For further customization, we can consider not showing the graph in the simulation for this group of students so that they can just focus on the table of values. More interactivity can be provided for the better students. In study of Park et al., (2009), it was reported that students with higher prior knowledge performed better with high-interactivity simulations while students with low prior knowledge reacted more positively with low-interactivity simulations.

## Conclusion

In this study, the energy simulation was adapted and implemented in an elementary school. We analyzed the findings from both the teachers' and students' perspectives. Due to the exploratory approach, the findings might not be generalizable but can be transferred to similar context. In this study, we had surfaced the generic principles with its corresponding TPACK construct in the design and customization stages. These would be useful to interested educators and researcher who wish to adapt and use simulation. For teacher-educators, the TPACK construct that emerged from the stages could provide some insights on designing lessons on customizing simulations. The findings also revealed that the students enjoyed learning with simulation. The positive learning experience could be due to deliberate features put in (e.g., the indication height) and the self-directed learning afforded by the simulation (interactivity and experimentation).

## Acknowledgement

We wish to express our deepest gratitude to the following group of people or organizations for making this research possible:

1. Francisco Esquembre, Fu-Kwun Hwang and Wolfgang Christian for their contribution to the open source physics simulation.
2. Michael Gallis for creating and sharing his original simulation